\documentstyle [12pt] {article}

\catcode`@=12
\begin{document}
\thispagestyle{empty}
\begin{flushright} UCRHEP-T95\\June 1992\
\end{flushright}
\vspace{0.5in}
\begin{center}
{\Large \bf Fixing the Top-Quark and Higgs-Boson Masses\\}
\vspace{1.5in}
{\bf Ernest Ma\\}
\vspace{0.3in}
{\sl Department of Physics\\}
{\sl University of California\\}
{\sl Riverside, California 92521\\}
\vspace{1.5in}
\end{center}
\begin{abstract}\
If the quadratic divergence of the standard electroweak model and its local
variation with mass scale are both assumed to be zero, then a modified one-loop
calculation yields $m_t \simeq 117~GeV$ and $m_H \simeq 183~GeV$.
Such a scenario may be the result of an underlying theory to be revealed
at a much higher mass scale.
\end{abstract}
\newpage

	The standard $SU(2)~X~U(1)$ electroweak gauge model has a fundamental
scalar doublet $\Phi = (\phi^+,\phi^0)$, which necessarily gives rise to a
quadratic mass term $\mu^2 \Phi^\dagger \Phi$.  As a result, there exists
a quadratic divergence in the quantum field theory which must be absorbed
into its redefinition.  Since this requires very delicate fine tuning, it
is considered to be an "unnatural" feature of the standard model.  It may
also be a hint that the standard model is either incomplete or just an
effective theory, at least in its scalar sector.  The former idea is
implemented by making the theory supersymmetric, so that bosonic and
fermionic contributions to all quadratic divergences cancel exactly in the
symmetric limit.  The latter idea is usually implemented by assuming that
there are no fundamental scalar fields and the standard-model $\Phi$ is a
condensate of fundamental fermions with new strong interactions at some
higher mass scale $\Lambda$.  If $\Lambda$ is only one to two orders of
magnitude larger than the electroweak mass scale of $10^2~GeV$, then the
existence of the standard-model quadratic divergence is not really a
problem.  However, if $\Lambda$ turns out to be much larger, then in order
for such a scenario to still make sense, the quadratic divergence itself
must be suppressed by a relationship among the effective parameters of the
theory, namely the various {\em a priori} independent couplings.  In other
words, a kind of screening should have taken place.

	In the standard model, let the Higgs potential be written as
\begin{equation}
V~=~\mu^2 (\Phi^\dagger \Phi) + {1 \over 2} \lambda (\Phi^\dagger \Phi)^2,
\end{equation}
then the vacuum expectation value of $\Phi$ is
\begin{equation}
v = (-\mu^2/\lambda)^{1 \over 2},
\end{equation}
and the mass of the physical Higgs boson is given by
\begin{equation}
m_H^2 = 2 \lambda v^2.
\end{equation}
The masses of the vector gauge bosons are related to $v$ by
\begin{equation}
M_W^2 = {1 \over 2} g_2^2 v^2,
\end{equation}
and
\begin{equation}
M_Z^2 = {1 \over 2} (g_1^2 + g_2^2) v^2,
\end{equation}
with the weak mixing angle $\theta_W$ given by
\begin{equation}
\sin^2\theta_W = {g_1^2 \over {g_1^2 + g_2^2}}.
\end{equation}
The mass of a given fermion $f$ is then
\begin{equation}
m_f = g_f v.
\end{equation}
The quadratic divergence of the $\mu^2$ term, calculated to one loop, is
proportional to a sum of squares of couplings weighted by certain numerical
factors which are positive for bosons and negative for fermions.  Let it be
set equal to zero.  Then
\begin{equation}
3 g_t^2 = {3 \over 8} g_1^2 + {9 \over 8} g_2^2 + {3 \over 2} \lambda,
\end{equation}
where all other Yukawa couplings ({\em i.e.} $g_b$, $g_\tau$, etc.) have
been neglected.  If the above equation is multiplied on both sides by $v^2$,
it becomes
\begin{equation}
3 m_t^2 = {3 \over 4} M_Z^2 + {3 \over 2} M_W^2 + {3 \over 4} m_H^2,
\end{equation}
which is a well-known condition first given by Veltman.\cite{1}$~$  Note that
Eq.(8) is unchanged whether one considers spontaneous symmetry breaking or
not.

	If Eq.(9) is valid, then the top-quark and Higgs-boson masses are
simply related.  In fact, it is consistent with the present
experimental data $M_Z = 91.175 \pm 0.021~GeV$,\cite{2}  $M_W = 80.14 \pm
0.27~GeV$,\cite{3}$~$  $m_t > 91~GeV$,\cite{4}$~$  and $m_H > 59~GeV$.\cite{5}
$~$  However, one cannot fix $m_t$ and $m_H$ separately.  The question now
is whether the standard model offers another hint for a second independent
condition relating them.  For that purpose, note first that Eq.(8) is
actually ambiguous because the values of the couplings are really functions
of mass scale.  In the standard model, this is naturally taken to be the
electroweak mass scale of about $10^2~GeV$, and Eq.(9) is well-defined.
On the other hand, it has been assumed that the mass scale $\Lambda$ for
new physics is much higher than $10^2~GeV$, hence a possible condition
is to require that the deviation of Eq.(8) away from $10^2~GeV$ be as
small as possible.  In other words, the local variation of Eq.(8) with
respect to a change in mass scale may be assumed zero in the belief that
$\Lambda$ is well screened.

	Let $t$ be minus the logarithm of the scale by which the
renormalization point is changed.  Then in the standard model, the relevant
one-loop renormalization-group equations are given by
\begin{equation}
8 \pi^2 { {dg_1^2} \over {dt} } = \left( {20 \over 9} N + {1 \over 6} \right)
g_1^4,
\end{equation}
\begin{equation}
8 \pi^2 { {dg_2^2} \over {dt} } = \left( {4 \over 3} N - {43 \over 6} \right)
g_2^4,
\end{equation}
\begin{equation}
8 \pi^2 { {dg_t^2} \over {dt} } = g_t^2 \left( {9 \over 2} g_t^2 - {17 \over
12} g_1^2 - {9 \over 4} g_2^2 - 8 g_3^2 \right),
\end{equation}
and
\begin{eqnarray}
8 \pi^2 { {d\lambda} \over {dt} } = 6 \lambda^2 - \left( {3 \over 2} g_1^2
+ {9 \over 2} g_2^2 - 6 g_t^2 \right) \lambda + {3 \over 8} g_1^4 +
{3 \over 4} g_1^2 g_2^2 + {9 \over 8} g_2^4 - 6 g_t^4,
\end{eqnarray}
where $N$ denotes the number of generations of quarks and leptons and will be
set equal to 3 in the following.  Let both sides of Eq.(8) be differentiated
with respect to $t$ according to Eqs.(10) - (13).  Now use it again to
eliminate $\lambda$.  The resulting condition is then given by
\begin{equation}
{21 \over 2} g_t^4 - \left( {23 \over 6} g_1^2 + {27 \over 2} g_2^2 - 8 g_3^2
\right) g_t^2 + {17 \over 12} g_1^4 + {21 \over 8} g_1^2 g_2^2 + {11 \over 4}
g_2^4 = 0.
\end{equation}
If Eq.(8) were valid for all mass scales, then the above
would be automatically zero and the theory would be guaranteed by Eq.(8)
alone to be free of quadratic divergences.  Indeed, this procedure was used
already several years ago\cite{6} to investigate whether there are
theories other than those with supersymmetry which might be free of quadratic
divergences.  Another procedure was to calculate the quadratic divergence
to two loops.\cite{7}$~$  If the vanishing of the one-loop contribution
automatically guarantees the vanishing of the two-loop contribution, then
the theory may well be free of quadratic divergences.  However, no example
of a nonsupersymmetric theory was found in Refs.[6] and [7].  It was then
pointed out two years ago\cite{8} that if a certain condition is
satisfied, then the vanishing of the one-loop contribution
will indeed lead to that of the two-loop contribution
automatically, but it remains unclear as to whether there are solutions
satisfying this condition outside of supersymmetry.

	It was also mentioned in Ref.[8] that in the standard model, if the
two-loop contribution to the quadratic divergence is set to zero in
addition to Eq.(8), then there is in fact no solution for $m_t$ and $m_H$.
However, this only tells us that the standard model cannot be like a
supersymmetric theory where all loop contributions to quadratic divergences
are zero.  What is required here is only the vanishing of the sum of all
the loop contributions.  Hence the two-loop term should be considered as
a correction to the one-loop term, and there is still only just one condition
on the two parameters.  Perhaps it should also be noted that if dimensional
regularization is used to extract the one-loop contribution of the quadratic
divergence, there is a dependence on the space-time dimension $d$ in the
residue of the pole at $d=2$.  If both this $d$ and the Dirac trace are set
equal to 4, as was done in Ref.[1], Eq.(9) is obtained and this agrees with
the usual cutoff procedure as well as the point-splitting method proposed
recently.\cite{9}$~$  On the other hand, if both are set equal to 2, then
\begin{equation}
{3 \over 2} m_t^2 = {1 \over 4} M_Z^2 + {1 \over 2} M_W^2 + {3 \over 4} m_H^2
\end{equation}
is obtained instead.  In this case, setting both the one-loop and two-loop
terms to zero results in the solution\cite{10}$~$ $m_t \simeq 174~GeV$ and
$m_H \simeq 232~GeV$.

	Returning to Eq.(14), it can easily be shown that it has no solution
for positive $g_t^2$.  This means that either the assumption of zero local
variation is not valid or that the underlying dynamics is not really
governed by Eq.(14).  If the latter viewpoint is adopted, then the only
possibility is that the $g_3^2$ term should be dropped.  The physical
reasoning behind this is simple: mass is an electroweak phenomenon and the
underlying theory which allows Eq.(8) to be valid may have nothing to do with
color $SU(3)$.  Dropping the $g_3^2$ term in Eq.(14) and using Eqs.(5) - (7),
one then finds
\begin{equation}
\left( {m_t^2 \over M_Z^2} \right)^2 - \left( {18 \over 7} \cos^2\theta_W +
{46 \over 63} \sin^2\theta_W \right) \left( {m_t^2 \over M_Z^2} \right) +
{22 \over 21} \cos^4\theta_W + \sin^2\theta_W \cos^2\theta_W + {34 \over 63}
\sin^4\theta_W = 0.
\end{equation}
For $\sin^2\theta_W = 0.233$, there are two positive roots to the above
quadratic equation, but one is not acceptable because $m_H^2$ would then
be negative; the other is $m_t^2/M_Z^2 = 1.64$, hence
\begin{equation}
m_t \simeq 117~GeV,
\end{equation}
and
\begin{equation}
m_H \simeq 183~GeV
\end{equation}
are obtained.  These values are of course corrected by higher-order
contributions to both Eqs.(9) and (16), but will probably not change by
more than a few percent.  Once $m_t$ is measured, it will then be known
whether or not Eq.(17) is correct.  However, it should be kept in mind
that Eq.(9) may be valid even if Eq.(16) is not.  In that case $m_t$ is
not fixed, but is only related to $m_H$.  It is also interesting to note
that the predicted value for $m_H$ is essentially just $2M_Z$, suggestive of
the relation $\lambda = g_1^2 + g_2^2$.  This may
be just an accident, or it may be another hint for the underlying dynamics.
In Ref.[9], in addition to Eq.(9), the condition that the $He\overline{e}$
coupling be finite is assumed, resulting in the prediction $m_t \simeq 120~
GeV$ and $m_H \simeq 190~GeV$.  However, from the viewpoint being advocated
in this paper, it is not clear why this particular coupling is to be singled
out.

	Other ideas for fixing $m_t$ and $m_H$ have appeared in the
literature.  One method\cite{11} is to match $g_t^2$ and $\lambda$ with $g_3^2$
and require that their ratios be unchanged with mass scale in the absence
of all the other couplings.  The contributions of $g_1^2$ and $g_2^2$ are
then treated as corrections.  This seems to imply that $m_t$
and $m_H$ may be closely related to color $SU(3)$ which is not at all
obvious.  The most recent analysis\cite{12} made the predictions $m_t = 99.2
\pm 5.7~GeV$ and $m_H = 64.6 \pm 0.9~GeV$, whereas an earlier independent
analysis\cite{13} had $m_t \simeq 95~GeV$ and $m_H \simeq 73~GeV$.  Another
method\cite{14} is to match $\lambda$ with $g_1^2$ which is expected to hold
at some very large mass scale and then use the renormalization-group
equations involving all the other couplings to extrapolate back to the
electroweak mass scale.  This allows $m_H$ to be determined as a function
of all the other masses, including $m_t$.  To fix $m_t$ as well, one may
assume that it should take on its maximum value.  Procedurally, this turns
out to be equivalent to the analysis of a recent model of dynamical symmetry
breaking.\cite{15}$~$  For a given $\Lambda$, the same values of $m_t$ and
$m_H$ are obtained by both procedures.  In the matching of couplings, they
are limiting values from below; whereas in Ref.[15], they are limiting values
from above.  The reason is that they involve the same set of differential
equations.  These critical values of $m_t$ and $m_H$ were known long ago
\cite{16} and have been proposed as a probe\cite{17} of possible new physics.
For $\Lambda = 10^{15}~GeV$ for example, the predictions are\cite{15}
$m_t \simeq 229~GeV$ and $m_H \simeq 256~GeV$.

	In conclusion, it has been suggested in this note that the standard
electroweak model itself may contain hints of an underlying dynamics whereby
$m_t$ and $m_H$ can be fixed.  It is assumed that the standard-model
quadratic divergence and its local variation with mass scale are both
vanishing.  Using in addition the observation that the masses in question
are of electroweak origin, the color $SU(3)$ coupling is discarded from the
one-loop contribution to $g_t$.  The resulting two equations have a unique
solution, namely $m_t \simeq 117~GeV$ and $m_H \simeq 183~GeV \simeq 2M_Z$.
A brief review of predictions based on other ideas has also been included
for comparison.
\vspace{0.3in}
\begin{center} {ACKNOWLEDGEMENT}
\end{center}
\vspace{0.1in}
	I thank G. Moultaka for discussion and correspondence.
This work was supported in part by the U. S. Department of Energy
under Contract No. DE-AT03-87ER40327.

\newpage
\bibliographystyle{unsrt}

\end{document}